\documentclass[showpacs,preprintnumbers,floatfix]{revtex4}
\usepackage{graphicx}

\def\la{\langle }
\def\ra{\rangle }

\def\be{\begin{equation}}
\def\ee{\end{equation}}

\def\bea{\begin{eqnarray}}
\def\eea{\end{eqnarray}}

\begin{document}


\title{Anomalies in the low CMB multipoles and extended
foregrounds}

\author{L. Raul Abramo}
\email{abramo@fma.if.usp.br}
\affiliation{Instituto de F\'{\i}sica, 
Universidade de S\~ao Paulo \\
CP 66318, CEP 05315-970 S\~ao Paulo, Brazil}

\author{Laerte Sodr\'e Jr.}
\email{laerte@astro.iag.usp.br}
\affiliation{Departamento de Astronomia, Instituto Astron\^omico e Geof\'{\i}sico, 
Universidade de S\~ao Paulo \\
Rua do Mat\~ao, 1226 -- 05508-090 S\~ao Paulo, Brazil}

\author{Carlos Alexandre Wuensche}
\email{alex@das.inpe.br}
\affiliation{Divis\~ao de Astrof\'{\i}sica,
Instituto Nacional de Pesquisas Espaciais\\
Av. dos Astronautas, 1.758, CEP 12227-010, S\~ao Jos\'e dos Campos, Brazil}

\date{\today} 

\begin{abstract} 
We discuss how an extended foreground of the cosmic microwave
background (CMB) can account 
for the anomalies in the low multipoles of the CMB anisotropies.
The distortion needed to account for the anomalies is consistent with a 
cold spot with the spatial geometry of the Local Supercluster (LSC) and 
a temperature quadrupole of order $\Delta T_2^2 \approx 50 \, \mu$K$^2$.
If this hypothetic foreground
is subtracted from the CMB data, the 
amplitude of the quadrupole ($\ell=2$) is 
substantially increased, and the statistically improbable 
alignment of the quadrupole with the octopole ($\ell=3$) 
is substantially weakened,
increasing dramatically the likelihood of the ``cleaned'' maps.
By placing the foreground on random locations and then
computing the likelihood of the cleaned maps we can estimate the
most likely place for this foreground.
Although the 1-year WMAP data clearly points the location of
this hypothetical foreground to the LSC or its specular image
(i.e., the vicinity of the poles of the cosmic dipole axis),
the three-year data seems to point to these locations as well as 
the north ecliptic pole. 
We show that this is consistent with the symmetries of the 
cosmic quadrupole. We also discuss a possible mechanism that 
could have generated this foreground: the
thermal Sunyaev-Zeldovich effect caused by hot electrons in the LSC.
We argue that the temperature and density of the hot gas 
which are necessary to generate such an effect, though 
in the upper end of the expected range of values,
are marginally consistent with 
present observations of the X-ray background of spectral 
distortions of the CMB.
\end{abstract}


\maketitle

\section{Introduction}

The cosmic microwave background (CMB) anisotropies have been measured
with very high accuracy by WMAP \cite{WMAP3y,WMAP1}. 
Such a barrage of new data
seldom brings only confirmation of known theories and mechanisms, 
and WMAP is no exception: lack of higher correlations \cite{Correl}
and some curious correlations between large-scale anisotropies
are some of the most intriguing questions that have 
been raised by the WMAP data. In particular, two
problems have been pointed with present CMB observations, using
a wide variety of data, methods, maps and sky-cuts: 
first, that the quadrupole ($\ell=2$) has a lower-than-expected 
signal \cite{OTZH,Efstathiou,Wagg}.
Second, that the quadrupole and octopole ($\ell=3$) 
present an unexpectedly high degree of alignment
\cite{OTZH,Copi04,Schwarz04,LM05,Hansen04,Eriksen04a,OT06,Copi06}.
The combined statistics of these effects implies that our
CMB sky is only within the $\sim$ 0,01\% of randomly-generated maps
with such anomalous quadrupole and octopole.

It is important to note, first, that these large-angle
anomalies were already present in the COBE data \cite{COBE}, 
and they were confirmed by WMAP both in the 1-year 
and in the 3-year data releases
\cite{WMAP3y,WMAP1}. Second, even though the amplitude of
the octopole $C_3$ has increased in the three-year compared 
to the first-year WMAP data, 
the statistical relevance of the deviant multipoles $\ell=2$ and 
$\ell=3$ has remained
practically unchanged in the newly released three-year WMAP data 
\cite{WMAP3y}, whereas the outliers of the first-year data \cite{WMAP1} 
around $\ell \approx 20$, $\ell \approx 40$ and $\ell \approx 200$ 
have either disappeared or become much less of a source of concern
in the 3-year data.

These large-angle anomalies have motivated many ingenious
explanations, such as compact topologies \cite{UzanEllis03,Luminet,MGRT04}, a broken or
supressed spectrum at large scales \cite{Broken}
and oscillations superimposed on the primordial spectrum of density
fluctuations \cite{Jerome}.
When the low value of the quadrupole is combined 
with the alignment of the directions defined by these two
multipoles (their "normal planes" -- see \cite{Copi04,OTZH,Schwarz04,KW04,LM05,OT06,Copi06}), 
the overall chance of such a statistical fluctuation
is approximately 0.005\% - 0.02\% depending on the map and on the
mask -- i.e., only about one in 10000 randomly 
generated models have a lower $C_2$ {\it and} a more aligned quadrupole 
and octopole than the observed CMB sky.

As first noted by de Oliveira-Costa {\it et al.} \cite{OTZH}, the 
directions preferred by the quadrupole and the octopole 
point roughly towards the Virgo cluster --- 
which is in the general vicinity of the dipole and the equinox,
and has been dubbed the ``axis of evil'' \cite{LM05}.
These large-scale anisotropies appear when one compares the
northern/western galactic hemisphere (where 
Virgo and most of the Local Supercluster lie) with the 
southern/eastern hemisphere \cite{Correl} as well. 
Unusually high correlations
with the ecliptic have also been reported \cite{Schwarz04}.

In this paper we investigate whether these large-scale anomalies 
can be due to extended (large-angle) foregrounds that have so far
escaped detection.
In particular, we re-examine a speculation by Abramo
\& Sodr\'e \cite{AS} that the explanation for the observed properties 
of the quadrupole and octopole is a diffuse, large-angle CMB foreground 
spatially correlated with the region of the sky occupied by the 
local supercluster (LSC) -- which is a spot roughly $50^o \times 30^o$
centered around Virgo, at $(l,b)=(284^o,74^o)$.
The peak temperature of this foreground would have to be of
order $\sim 20 \mu$K, with a root-mean-square average temperature
of order $\sim 8 \mu$K.
Such a foreground would have the correct geometry to affect the
quadrupole and octopole in a positive way, erasing their alignments 
and significantly increasing the amplitude of the quadrupole.
One of the physical processes that could produce such a foreground
is the thermal Sunyaev-Zeldovich effect (SZe) due to hot electrons 
in the intra-supercluster (ISC) medium \cite{AS}. For the range of 
frequencies observed by WMAP and COBE, the ISC gas causes
an apparent decrease in the 
temperature of the CMB photons in the direction of the LSC.
We have estimated, using a simplified model, 
that a temperature distortion as high as
$|\Delta \hat{T}_{\ell=2}|_{rms} \approx 8$ $\mu$K is
marginally consistent with present constraints on the ISC medium
and with spectral distortions of the CMB.

The possibility that foregrounds could be responsible for the
alignments was also noted in \cite{Uzan05}, and
two recently related explanations were proposed by Rakic, Rasanen 
and Schwarz \cite{ReesSciama} and by Inoue and Silk \cite{Disk06}.
Rakic {\it et} al. studied the Rees-Sciama effect due to a nearby
large-scale structure (or structures), and found that the ensuing 
foreground could be as high as $\sim 30 \mu$K, but that the phases were not
right to eliminate the alignments and the low quadrupole.

Inoue and Silk, on the other hand, speculate that the
non-gaussian cold spot in the southern galactic hemisphere
\cite{Vielva} could be caused by voids in the nearby 
large-scale structure \cite{Disk06}. They conclude that the contribution from
compensated pairs of voids would have the right phases to 
account for the low amplitude of the quadrupole and the
for the unusual quadrupole-octopole alignment.

Notice that, because the quadrupole is even under parity transformations
$\hat{n} \rightarrow -\hat{n}$, any given pattern has the same quadrupole 
components as its specular image. Since the effect discussed here relies 
mostly on a distortion of the quadrupole
(which is both low-amplitude and has fewer phases than the octopole), 
this partly explains the apparent equivalence between 
the foreground proposed by Abramo \& Sodr\'e and foregrounds located
in the southern galactic hemisphere, such as those proposed by Inoue \& Silk.
Furthermore, because most of the power of the observed CMB quadrupole 
lies in its $m=0$ and $m=\pm 2$ components, another probable 
spot for an extended foreground corresponds to the location 
of Virgo, rotated $180^o$ around the galactic poles axis -- that is,
$(l,b) \approx (100^o,70^o)$.

By placing one of these hypothetical foregrounds (henceforth HFg)
on random locations, removing 
it from a CMB map and 
then computing the likelihood of the ``cleaned" map, we have been 
able to test the randomness of the spatial correlation with the 
four dual points described above.
We find that, in all maps and in all foreground models tested,
the most probable places for them
are indeed either the vicinities of $\hat{n}_{LSC}$, $-\hat{n}_{LSC}$
or near the ecliptic poles.

We will show that, of all possible locations for this 
HFg, the LSC and its dual points produce the most
significant improvements in the likelihoods of the CMB maps,
by both increasing the (too low) level of the quadrupole $C_2$ and by
weakening the (too high) quadrupole-octopole alignment.
This can be achieved with HFg's
whose rms temperatures 
lie in the range $5-15 \mu$K and quadrupoles 
$C_2^{Fgrd} \sim 50 - 120 \mu$K$^2$, depending on the model.

We have analysed the WMAP (1-year and 3-year) Internal Linear Combination maps
\cite{WMAP3y,WMAP1} (henceforth ILC), the map of Tegmark 
{\it et} al. \cite{TOH,OT06} (henceforth TOH), 
as well as the co-added
maps based on 1-year and 3-year WMAP data (henceforth Coadded)
For the ILC and Coadded maps we use the Kp2 maks, and for the TOH map we
use the masks M0 and M6 described in \cite{TOH}.
In all cases the low value of the quadrupole and the alignments are robust,
and removal of the HFg
leads to dramatic increases in the likelihoods of the CMB maps.
This strongly argues in favour of still unknown diffuse, large-angle 
structures around the dipole axis that may be affecting the CMB.

This paper is organized as follows: In Sec. II we summarize the
multipole vector formalism and the statistics of alignments for CMB
maps. In Sec. III we present a model of the 
HFg
based on the LSC, and argue that it may be due to the
Sunyaev-Zeldovich effect caused by hot gas in the intra-supercluster 
medium. In Sec. IV we consider the spatial location of the foreground, 
and show that the association with the dipole axis is not an accident.
We conclude in Sec. V.


\section{Low quadrupole and alignments}

Katz and Weeks \cite{KW04} have described, in a beautiful paper, how to
compute all multipole vectors given the spherical harmonic 
components $a_{\ell m}$ -- see also \cite{Germans}.
The multipole vectors, introduced to CMB data analysis by Copi {\it et al.}
\cite{Copi04}, are essentially eigenvectors
of a simple set of algebraic equation whose parameters are the multipole components.
Very similar computations were conducted using other (usually numeric) methods 
in \cite{OTZH,Correl,Schwarz04} to find these vectors. 
The idea, which goes back to J. C. Maxwell in the XIX$^{\rm th}$ century,
is that the multipole decomposition of a field on $S^2$ 
implies that for each moment $\ell$ there are $\ell$ eigenvectors of norm unity,
$\hat{n}^{(\ell,p)}$.
The bottom line of the multipole vector analysis is that the expansion in
spherical harmonics is equivalent to an expansion in multipole vectors:
\be
\label{expansions}
\frac{\Delta T_\ell(\theta,\varphi)}{T} = 
\sum_{m=-\ell}^\ell a_{\ell m} Y_{\ell m} (\theta,\varphi) 
= D_\ell \prod_{p=1}^\ell \hat{n}^{(\ell,p)} \cdot \hat{n} (\theta,\phi) - 
Z_{\ell-1}(\theta,\varphi) \; ,
\ee
where $Z_{\ell-1}$ just subtracts the residual $\ell ' < \ell$ total
angular momentum parts of the product expansion, and is irrelevant to
our analysis -- see \cite{KW04} for an enhanced discussion of the
multipole vector expansion.

Notice that, contrary to the $C_\ell$'s, which are always positive-definite,
the $D_\ell$'s can be either negative or positive. This means
that the multipole vectors $\hat{n}^{(\ell,p)}$ define only 
directions \cite{KW04}, hence they are in fact ``vectors without arrowheads".
It can also be seen from the expansion above that, whenever using the
multipole vectors to test for alignments, it is irrelevant what the
amplitudes of the multipoles are -- just their (complex) 
phases matter, of which there are $\ell$ for each multipole.

Starting with these $\ell$ multipole vectors one can also construct $\ell(\ell-1)/2$
normal vectors -- or normal planes. Therefore, for $\ell=2$ there are 2 multipole vectors
($\hat{n}^{(2,1)}$ and $\hat{n}^{(2,2)}$) and only one normal plane 
($\vec{w}^{(2,1)}= \hat{n}^{(2,1)} \times \hat{n}^{(2,2)}$); 
for $\ell=3$ there are 3 multipole vectors and 3 normal
planes; and so forth. Notice that, because the multipole vectors are not
necessarily orthogonal, the normal vectors need not be (and generally are 
not) of norm unity.

We can therefore check for ``alignments" between either the 
multipole vector themselves, or between the normal planes. Two widely used
tests that check for alignments of the quadrupole and octopole normal planes
are the $S$ statistic:
\be
\label{def:S}
S \equiv \frac13 | \vec{w}^{(2,1)} \cdot \vec{w}^{(3,1)} | + 
| \vec{w}^{(2,1)} \cdot \vec{w}^{(3,2)} |
+ | \vec{w}^{(2,1)} \cdot \vec{w}^{(3,3)} | \; ,
\ee
and the $D$ statistic, which is analogous to $S$ but 
disregards the norm of the normal vectors:
\be
\label{def:D}
D \equiv \frac13 | \hat{w}^{(2,1)} \cdot \hat{w}^{(3,1)} |
+ | \hat{w}^{(2,1)} \cdot \hat{w}^{(3,2)} |
+ | \hat{w}^{(2,1)} \cdot \hat{w}^{(3,3)} | \; .
\ee
It can be easily seen that both $S$ and $D$ lie within the interval $(0,1)$.
In what follows we will use mostly the $S$ statistic, since the $D$ statistic disregards the norm of the normal vector and therefore throws away some information about the system. For the 
higher multipoles this may not be much of an issue, but the 
quadrupole has only 2 complex phases (4 effective degrees of 
freedom) and we would like to retain as much of that phase 
information as possible.

\begin{figure}
\includegraphics[width=8cm]{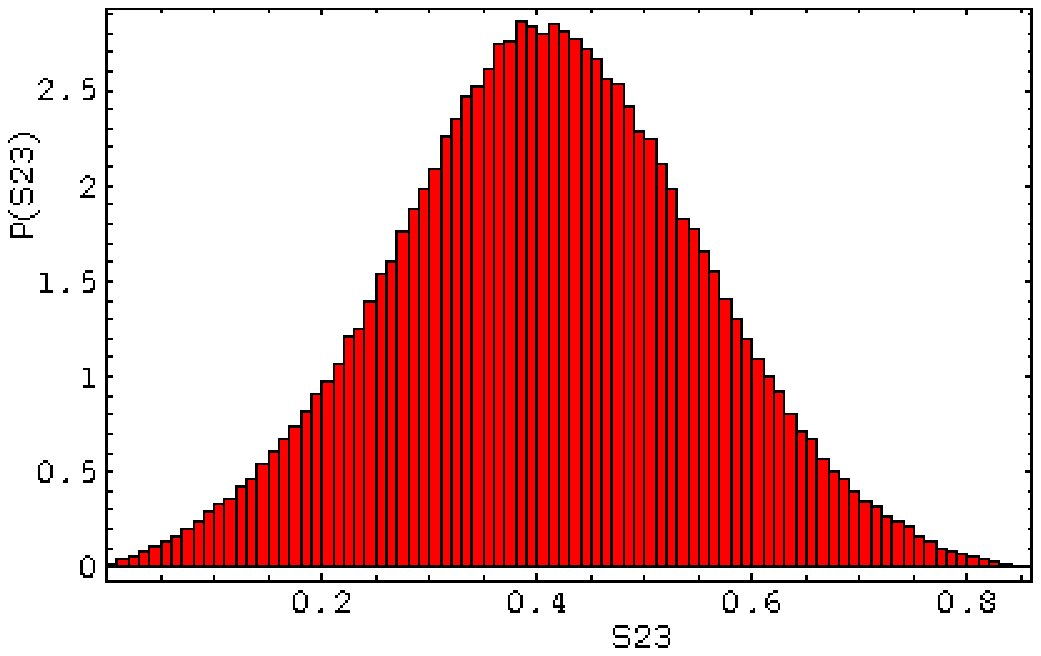}
\includegraphics[width=8cm]{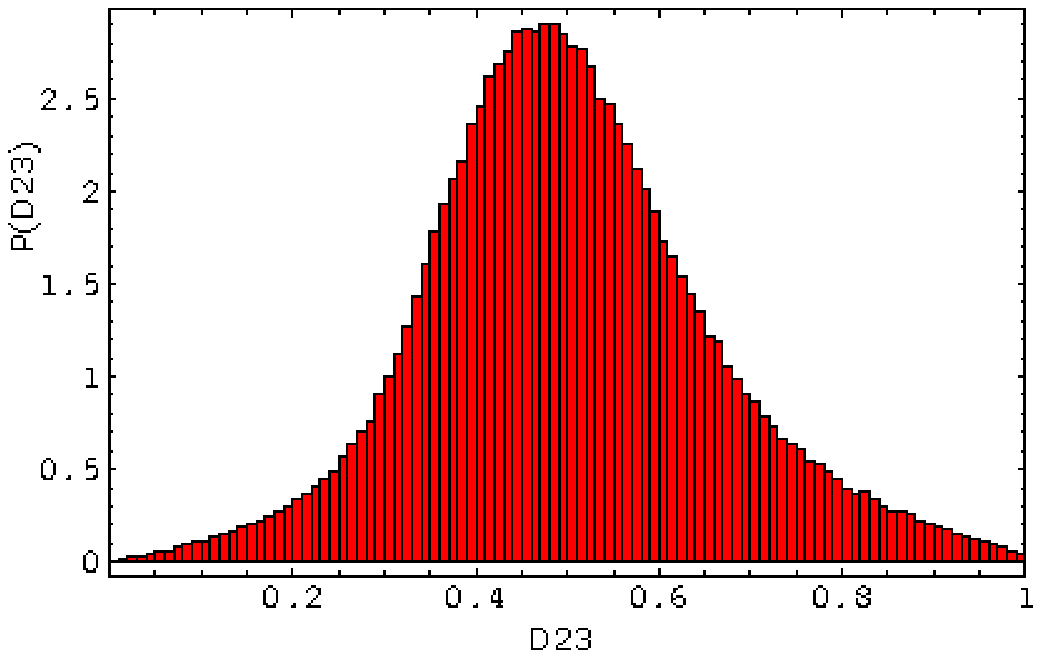}
\caption{\label{fig:histograms} Normalized P.D.F.'s for the $S$ statistic 
(left panel) and $D$ statistic (right panel), found by simulating 
$3\times 10^5$ mock maps, binned in intervals of 0.01.}
\end{figure}

One can easily compute the probability distribution functions (P.D.F.'s) for 
$S$ and $D$ using randomly-generated (``mock'') maps, either by simulating 
maps, computing the harmonic components and then the vectors and 
alignments, or by directly simulating the harmonic components, which are 
Gaussian random numbers with dispersion given by $\langle |a_{\ell m}|^2 
\rangle = C_\ell$. Because the alignments do not depend on the $C_{\ell}$'s, but 
only on the phases between the $a_{\ell m}$'s, one need not adjust the level of 
the $C_{\ell}$ for each $\ell$ -- all that is needed to test for the 
alignments in mock maps is the fact that the phases are random within 
each multipole.

In Fig. 1 we show the normalized P.D.F.'s for the $S$ and $D$ tests that 
were computed using 300.000 mock maps.
As mentioned above, in what follows 
we will use mostly the $S$ statistic, but we note that the $D$ statistic 
gives similar results. For a complete treatment of statistical tests of 
isotropy, their P.D.F.'s, and an assessment of other sources of error 
see \cite{Nois}.

\subsection*{Statistics of large-angle anisotropies and the low quadrupole}

Given a CMB map, the harmonic components can be easily extracted (we use
HEALPix \cite{HEALPix}), and the multipole vectors and their statistics
can be easily computed. There are several maps to choose from, the most
well-known being the Coadded, ILC, LILC, TOH and the Q-, V- and W-band 
frequency maps.
For all except the TOH map (which is already cleaned) we have use the KP2 mask
based on three-year WMAP data \cite{WMAP3y}.

It must be noted that the relativistic Doppler correction to the quadrupole 
is an important factor that must be subtracted from the maps, 
since it corresponds to a non-primary source of the quadrupole \cite{Schwarz04}.

In Table I we present the quadrupoles and their alignments with the
octopoles, for a few CMB maps. The Coadded and ILC maps use 3-year WMAP 
data, while the
TOH map uses the first-year WMAP data only. The alignments are robust
in all maps, as has been noted by \cite{Schwarz04,LM05,OT06}.
In the next section we will construct a model foreground based on
the LSC, and Table I presents the statistical effect of the 
subtraction of this LSC-shaped foreground. Notice that the 
relative error for the probabilities can be estimated as
$\Delta P(X)/P(X) \sim 1/ \sqrt{300.000 \times P(X)}$.

\begin{table}
\begin{tabular}{|c|c|c|c|c|c|c|}
\hline
CMB Map & 
$\; C_2$ ($\mu$K$^2$) $\;$ & $ \quad P_{-}(C_2)  \quad$ & $\quad S \quad $ & $ \;  \quad P_{+}(S)  \quad \;$ & $ \quad  \quad P_{\rm Tot}  \quad  \quad$  \\
\hline
\hline
TOH$_{\rm Mask \; 0}$ & 
201.2 & 3.43$\times 10^{-2}$ & 0.797 & 2.04$\times 10^{-3}$ & 1.6$\times 10^{-3}$ \\
TOH$_{\rm Mask \; 0}$ - LSC & 
340.6 & 1.02$\times 10^{-1}$ & 0.537 & 1.99$\times 10^{-1}$ & 2.5$\times 10^{-1}$ \\
TOH$_{\rm Mask \; 6}$ & 
242.0 & 5.09$\times 10^{-2}$ & 0.776 & 4.04$\times 10^{-3}$ & 4.1$\times 10^{-3}$ \\
TOH$_{\rm Mask \; 6}$ - LSC & 
399.7 & 1.38$\times 10^{-1}$ & 0.531 & 2.19$\times 10^{-1}$ & 3.3$\times 10^{-1}$ \\
Coadded (1yr) & 
97.7 & 6.69$\times 10^{-3}$ &  0.748 & 1.15$\times 10^{-2}$ & 1.2$\times 10^{-3}$ \\
Coadded (1yr) - LSC & 
222.3 & 4.25$\times 10^{-2}$ &  0.509 & 2.85$\times 10^{-1}$ & 1.3$\times 10^{-1}$ \\
Coadded (3yr) & 
100.5 & 7.16$\times 10^{-3}$ &  0.746 & 1.15$\times 10^{-2}$ & 1.3$\times 10^{-3}$ \\
Coadded (3yr) - LSC & 
174.7 & 2.52$\times 10^{-2}$ & 0.590 & 1.17$\times 10^{-1}$ & 4.1$\times 10^{-2}$ \\
ILC (1yr) & 
139.7 & 1.53$\times 10^{-2}$ & 0.727 & 1.74$\times 10^{-2}$ & 4.1$\times 10^{-3}$ \\
ILC (1yr) - LSC & 
277.6 & 6.77$\times 10^{-2}$ & 0.537 & 2.19$\times 10^{-1}$ & 1.7$\times 10^{-1}$ \\
ILC (3yr) & 
111.7 & 9.15$\times 10^{-3}$ & 0.720 & 2.10$\times 10^{-2}$ & 3.0$\times 10^{-3}$ \\
ILC (3yr) - LSC & 
207.8 & 3.68$\times 10^{-2}$ & 0.538 & 2.19$\times 10^{-1}$ & 9.7$\times 10^{-2}$ \\
\hline
\hline
\end{tabular} 
\caption{\label{Table1} Quadrupoles and alignments of CMB maps  
with and without the hypothetical LSC foreground subtracted. 
Shown are 
the Coadded and ILC maps with the 3-year KP2 mask \cite{WMAP3y,WMAP1}, and
the TOH map with and without the mask described in \cite{OT06} (based on 
first year WMAP data). $P_-(C_2)$ is the probability that a random map has
quadrupole as low as $C_2$, and $P_+(S)$ is the probability that a
random map has a quadrupole-octopole alignment as high as $S$.
Also shown (last column) is the unbiased joint 
probability
$P_{\rm Tot} = 16 \times P_- (C_2) \times P_+ (C_2) \times P_- (S) \times P_+(S)$,
which estimates the likelihood that a random map
has an anomalous (too high or too low) quadrupole {\it and} an anomalous
(too high or too low) quadrupole-octopole alignment. 
In all cases, removal of the foreground leads to an improvement of about 
two orders of magnitude in $P_{\rm Tot}$.}
\end{table}

\section{Hypothetical foreground}

As first noted by de Oliveira-Costa {\it et} al. \cite{OTZH,OT06}, both
the quadrupole and the octopole seem to be aligned on the plane
defined by the direction $(l,b) \approx (250^o,60^o)$, which is
quite close to the Virgo cluster. This motivated the proposal of
Abramo \& Sodr\'e \cite{AS}, who speculated that the low-$\ell$ anomalies
of the CMB could be explained by the Sunyaev-Zeldovich effect caused
by hot electrons in the intra-supercluster medium of the LSC. 
For the frequency channels
observed by COBE and WMAP, this effect
would cause a cold spot with the shape and location of the
supercluster superimposed on the primary CMB data.

We will show next that there are strong indications that an LSC-related
foreground (or some structure diametrically opposite to the LSC) is 
distorting the observed CMB sky. 
Whether or not that foreground is caused by the SZe \cite{AS}, 
a void \cite{Disk06}, some other 
mechanism such as the Rees-Sciama (or Integrated Sachs-Wolfe) effect 
\cite{ReesSciama}, or even a combination of those, remains to be seen.

\subsection{Shape and location of the Local Supercluster}

The morphology of the LSC is relatively well known \cite{LSCmorph}:
it is a flattened collection of groups and clouds of galaxies 
centered at the Virgo Cluster, which contains $\sim$20\% of 
its bright galaxies.
The Local Group is dynamically linked to the LSC, and lies
$\sim$15 Mpc from Virgo, at the border of the LSC.
Notice that the LSC itself is not a virialized structure,
hence the gas in its midst is not necessarily in equilibrium.

Since we are interested in an analytic approach at this point,
a radical simplification will be made, approximating
the shape of the LSC by an oblate spheroid of maximal 
radius 20 Mpc with approximate axial ratios 6:2:1 \cite{LSCmorph}.
Therefore, our simple model assumes that the LSC is a collection
of objects (clouds, groups and the Virgo cluster) which are
distributed smoothly across the spheroid.
The Sun stands at the margin of the spheroid 
(which looks like a flattened rugby ball), 
approximately 15 Mpc away from Virgo.

In our foreground model we assume that the intensity of the temperature
decrement is proportional to the volume of the LSC, projected along the line of
sight (i.e., the surface density.) 
This must be roughly correct, whatever the source of the hypothetical
foreground, if it is indeed correlated with a diffuse structure such as the 
LSC. It is trivial to compute the surface density, given the
shape and orientation of the LSC, and the result of this projection
can be seen in Fig. 2 for an arbitrary (but constant) density. 
In that figure it can also be seen that our LSC model's projection
on our sky is a spot of roughly $50^o \times 30^o$.

\begin{figure}
\includegraphics[width=12cm]{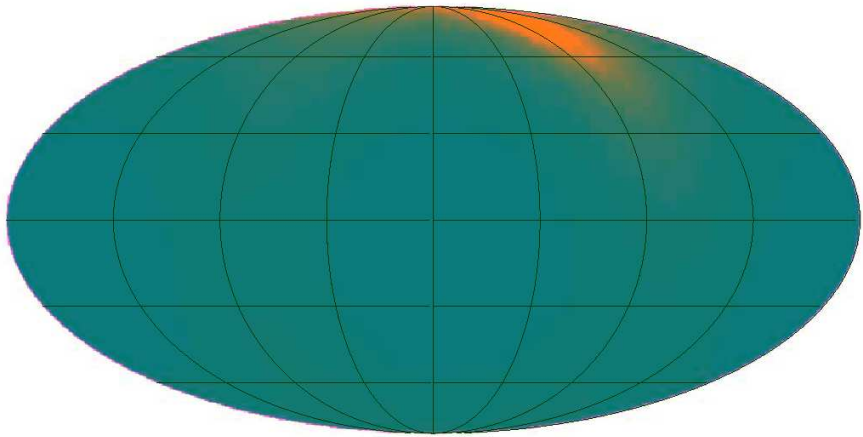}
\caption{\label{fig:LSC} Projected volume of the LSC on a Mollweide projection -- 
the left edge corresponding to $l=0^o$, the middle meridian to $l=180^o$ and
the horizontal line to $b=0^o$ in galactic coordinates. 
Virgo is at the center of the LSC, at $(l,b) \approx (284^o,74^o)$.}
\end{figure}

The oblate spheroid, in a conveniently rotated frame, is defined by the surface:
\be
\label{surface}
\left( b x' \right)^2 +
\left( c y' \right)^2 +
{z'}^2 = A^2
\; ,
\ee
where $A$ is the major axis (along the direction $z'$), 
and $b$ and $c$ are the ratios of the minor axes to the major axis.

The LSC parameters are $A \approx 20$ Mpc, $b \approx 3$ and $c \approx 6$.
With these values the semi-major axes are $B=A/b\approx 6.7$ Mpc and 
$C=A/c\approx 3.3$ Mpc.
Assuming that the Sun is located at a distance $R$ under the $z'$-axis of
the spheroid, the distance to the surface of the spheroid along 
lines-of-sight emanating from the Sun are given by:
\bea
\label{ra}
r_{S} (\theta',\phi') &=& \frac{1}{1+\sin^2\theta' 
\left[ (b^2-1) \cos^2\varphi' + (c^2-1) \sin^2\varphi' \right]}
\\ \nonumber
&\times& \left[ R \cos\theta' + 
\sqrt{R^2 \cos^2\theta' + (A^2 - R^2) \left[ 
1+\sin^2\theta' \left( (b^2-1)\cos^2\varphi' + 
(c^2-1) \sin^2\varphi' \right) \right] }
\right] \; .
\eea
Obviously, if the density is uniform inside the spheroid 
then the surface density will be proportional to $r_S$.

We can use the surface density of the LSC as the blueprint for a
foreground, and therefore consider the temperature decrement caused 
by the foreground to be proportional to $r_S$. 
In Fig. 3 (left panel) we show the spectrum of anisotropies
for such a model, where the proportionality constant is set by
assuming that the effect is caused by scattering of the CMB photons by
hot electrons in the ISC medium -- see below. 
Notice that the quadrupole is substantially higher than the other
multipoles, because the foreground's temperature is not constant as the
line of sight moves away from the center of the LSC. This sorts out
the quadrupole as the biggest contribution to the power spectrum,
and in fact the RMS temperature fluctuation over the whole sky
is well approximated by the quadrupole.

\begin{figure}
\includegraphics[width=8cm]{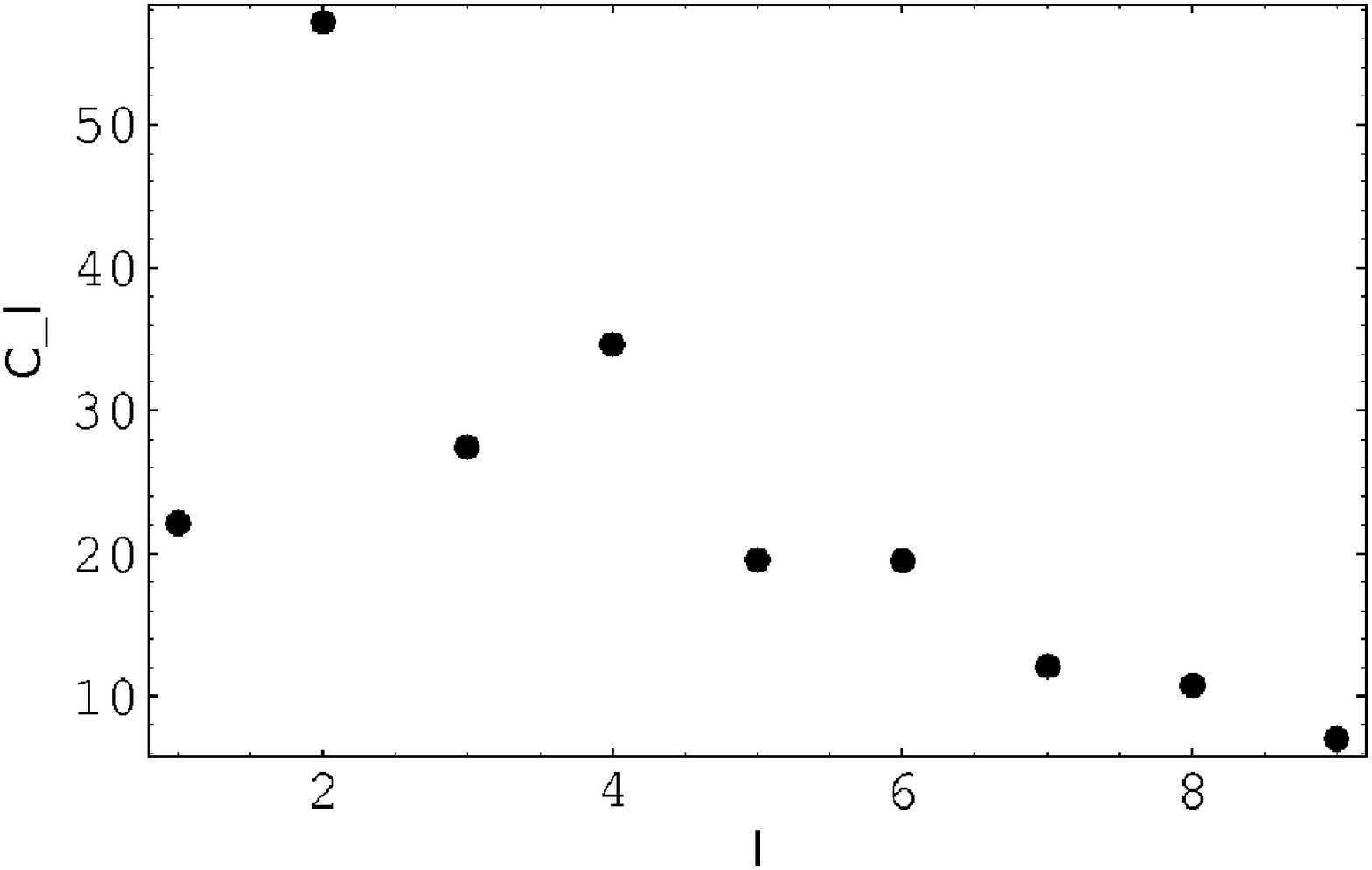}
\includegraphics[width=8cm]{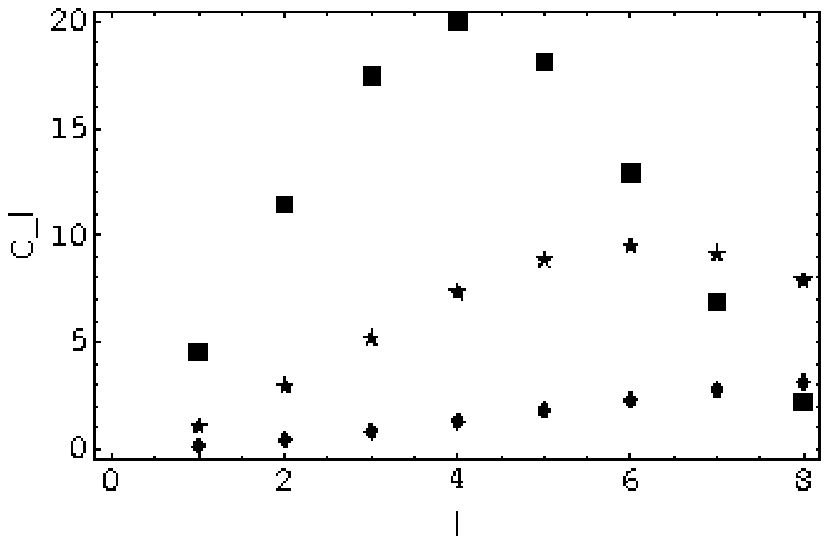}
\caption{\label{fig:3} Angular power spectra
of the LSC foreground model (left panel), where we use $\alpha=1$ -- see Eq. (\ref{quadrupole});
and of the Disk (uniform-temperature) foreground model, with angular diameters of
$30^o$ (diamonds), $50^o$ (stars) and $70^o$ (squares.)}
\end{figure}

Given our ignorance about the existence, shape and form of this 
hypothetical foreground,
we could equally well assume, following Inoue \& Silk \cite{Disk06},
that the temperature of the 
foreground is approximately uniform, falling quickly to zero away
from the center of the spot. In Fig. 3 (right panel) we show the anisotropy 
spectra of three such spots -- with angular diameters  of $30^o$, $50^o$ 
and $70^o$. 
The fact that the temperature is uniform inside the spot means that
the low multipoles contribute more evenly to the foreground. Hence, 
for a given RMS temperature fluctuation there is less power in the
quadrupole in the uniform-temperature foreground relative to the 
varying-temperature foreground. Moreover, as higher multipoles
are more important in the uniform-temperature foreground, if
that is the case then it may be possible to find independent 
corroborating evidence by searching for anomalous 
alignments in the higher multipoles (e.g., $\ell=4$ and $\ell=5$) 
as well. This will be analysed in Sec. IV.

\subsection{Foreground model: Sunyaev-Zeldovich effect in the LSC}

The SZe is caused by the inverse Compton scattering
of CMB photons by hot electrons in the intra-cluster medium \cite{SZ}. 
It is a nonthermal, frequency-dependent effect:
the upscattering causes an incident blackbody spectrum of photons to 
become distorted in such a way that the resulting 
higher abundance of high-energy photons 
is compensated by a shortage of low-energy photons. The 
spectral distortion is given by \cite{SZ}:

\be
\label{specSZZ}
\frac{\Delta T (\theta,\varphi;\nu)} {T_0} = 
y(\theta,\varphi) \left( x \, {\rm cotanh} \, \frac{x}{2} -4 \right) \; ,
\ee
where $T_0=2.726$K is the temperature of the CMB, 
$y$ is the comptonization parameter in the direction
$\hat{n}(\theta,\varphi)$ and $x=h\nu/k_{B}T_0$.

The frequency at which photons are neither depleted nor overproduced
is $\nu_0 = 218$ GHz \cite{RevSZ} --- COBE/DMR and WMAP work in the 
range 20-90 GHz. For frequencies below $\nu_0$
the effect is a nearly uniform reduction in the temperature of the 
photons, $\Delta T/T \approx - 2 y$, and
for frequencies above that the effect is the opposite.
This means that measurements
over a range of frequencies around $\nu_0$ (such as PLANCK's LFI and HFI
\cite{PLANCK}) can pick up the signal of the SZe and
distinguish it from the primary anisotropies.

The comptonization parameter $y$ measures an
optical depth for the CMB photons created by the hot electrons, 
and its value is given by the product of the Thomson cross-section 
$\sigma_T = 6.65 \times 10^{-25}$ cm$^2$
times the temperature-averaged density of photons along the line of sight
\cite{RevSZ}:
\be
\label{y}
y = \int \sigma_T \frac{k T_e}{m_e c^2} n_e dl \; ,
\ee
where $T_e$ is the electron temperature, $m_e$ is the electron mass
and $dl=dl(\theta,\phi)$ is the line-of-sight distance element along
the direction $(\theta,\phi)$.

The SZe has been observed over the past few years in many clusters, 
but its weak strength means that it could only be
detected in the central parts of clusters, where column densities
of hot gas are sufficiently high \cite{RevSZ,SZObs}.
It is evident that some amount of SZ will take place also
in the LSC, but the question is, how much? The answer depends on the 
gas density in the ISC medium, its temperature distribution, 
the morphology of the LSC and our position inside it.

Although the morphology of the LSC as traced by galaxies
is well known, the density and 
temperature distribution of the gas of the ISC medium are not.
Unfortunately, X-ray and microwave observations have not yet 
reached the level of sensitivity required
to detect directly the very smooth, diffuse columns of hot gas in 
the outer regions of clusters.
It seems, however, obvious that there must be a great amount
of ionized gas in the ISC medium, among other reasons because 
the absence of observations of the Gunn-Peterson effect indicates that
most of the ISC hydrogen must be ionized.
The gas is thought to have been shock-heated at the time of galaxy formation, and
now it is probably distributed in many phases, including filaments and 
a more homogeneous component \cite{CO,Kravtsov,Sembach}.
Phillips, Ostriker and Cen \cite{POC} have constrained the amount 
of gas in filaments 
using numerical simulations and 
the X-ray background, and argued that this ``warm-hot'' 
($kT \approx$ 100 eV -- 10 keV) gas can account for only 5--15 \% 
of the ``missing baryons''. More recently, Nicastro {\it et al.} showed
that this fraction could be as high as 27\% \cite{Nicastro05}. 
It is therefore quite possible that much 
of this gas is in the ISC medium.
So, the questions now are:
how hot is this ionized gas, and how is it distributed?

Hogan was the first to propose 
that superclusters (and the LSC) could 
impact the CMB anisotropies through the thermal and kinetic
Sunyaev-Zeldovich effects \cite{Hogan}.
Molnar and Birkinshaw used HEAO 1 A2 \cite{HEAO} and COBE DMR data to 
analyze the Shapley supercluster and found no evidence of hot 
($>10^7$K) gas in the ISC medium \cite{MB}.
Boughn \cite{Boughn}, on the other hand,
used the HEAO 1 A2 X-ray map and a simple ``pillbox'' model
of nearly constant electron density in the LSC to argue that the
SZe could be as high as 
$|\delta T| \sim (17 \pm 5)$ $\mu$K --- although he assumed
a gas temperature in the high end of the range $10^5$--$10^8$ K.
Kneissl {\it et al.} \cite{Kneissl} 
did study the correlation of COBE DMR and ROSAT X-ray
data away from the galactic plane, but it is not clear that the X-ray 
data has enough sensitivity to detect the diffuse hot gas of the LSC, 
and, in any case, the authors analyzed a region which 
misses a large chunk of the LSC.

Much work has been done to study the impact of the SZe from
{\it distant} clusters on the CMB (see, e.g. \cite{Diaferio,MB2,Uros,SR}).
It has been found that the
largest contribution to the angular power spectrum from the SZe
comes from the most massive clusters 
($M \sim 10^{15} h^{-1} M_\odot$), at scales $\ell \sim 3000$, with 
amplitudes
$\ell(\ell+1)C_\ell/2 \pi \approx 10 - 100 \, \mu$K$^2$.



The overall number of free electrons in the LSC can be estimated
given its gas fraction and mass:
\be
\label{Ne}
N_e = \frac{M_{LSC} f_g}{ \mu_e m_p} \; ,
\ee
where $M_{LSC}$ 
is the LSC mass, $f_g$ is the gas fraction, 
$\mu_e$ is the molecular weight per electron and 
$m_p$ is the proton mass. 
We may assume that the mass of the LSC is
$\sim 7 \times 10^{15}$ M$_\odot$ \cite{SPT}. 
Assuming that the Hydrogen is fully ionized
and that the helium mass fraction is
$Y=0.24$, then $\mu_e = 1/(1-Y/2) \simeq 1.14$. 

The gas fraction is not very well known, but X-ray observations of 
clusters indicate that $f_g \approx 0.06 h^{-3/2}$ \cite{EttoriFabian99}. 
The fraction could be different in the ISC medium, but we will assume
for simplicity that the fraction in clusters is the same as outside.
Using $h=0.7$ we get finally that the total number of electrons in the
LSC should be of order $N_e \sim 7 \times 10^{71}$ .

A very simple model for the LSC gas is to neglect the gas phases
and the substructures inside the supercluster, and to assume that 
the gas is uniformly distributed inside the LSC. Hence the total
volume occupied by the gas is $V_{LSC} = 4 \pi/3 \times A B C$
and therefore the average density of electrons in the LSC is approximately:
\be
\label{n_e}
n_e = \frac{N_e}{V_{LSC}} \sim 1.3 \times 10^{-5} {\rm cm}^{-3} \; .
\ee


On the other hand, the X-ray background is also an important 
constraint on the density and temperature of the ISC medium. 
A compilation of observations [20] gives a
background flux for energies $h \nu \sim 2$ keV of approximately $10^{-25}$
erg s$^{-1}$ cm$^{-2}$ sr$^{-1}$ Hz$^{-1}$ over the whole sky. 
The expected flux at this energy due to thermal 
bremsstrahlung emission from LSC gas 
is $\sim 5 \times 10^{-26}$ erg s$^{-1}$ cm$^{-2}$ sr$^{-1}$ Hz$^{-1}$.
Since the X-ray flux is proportional to the square of the electron density,
if the gas temperature is indeed 2 keV, then the upper bound for the
electronic density is of order $n_e \approx 5 \times 10^{-5}$ cm$^{-3}$.
This corresponds to a collapse factor of only a few.

We can estimate the comptonization parameter assuming a constant 
electron density across the LSC. If the gas has an {\it average} 
temperature of 2 keV then $\la k T_e \ra / m_e c^2 \simeq 0.004$, and
with a maximum line-of-sight distance (in the direction of Virgo)
of $\sim 30 \, {\rm Mpc}$ we obtain that
the comptonization parameter is at most:
\be
\label{comptoniz}
\Delta y \approx \sigma_T \la \frac{k T_e}{m_e c^2} \ra
\, \times n_e \times 30 {\rm Mpc} \sim \frac{n_e}{5. \, 10^{-5} {\rm cm}^{-3} } 
\times \, 10^{-5} \; .
\ee

The comptonization parameter can be exactly computed from Eq. (\ref{y})
for our oblate spheroid model, if we assume that the density
and temperature of the hot gas is uniform inside the LSC, and zero 
outside it. In that case the comptonization parameter is proportional
to the projected distance to the surface of the spheroid, Eq. (\ref{ra}).
The resulting angular power spectrum for the SZe of the LSC in this
model is given in Fig. 3. The amplitude of the SZe quadrupole is:

\bea
\label{quadrupole}
\Delta \hat{T}_2^2 &\equiv& \frac{6}{2 \pi} \hat{C}_2 
\approx 60 \, \alpha^2 \, \mu{\rm K}^2 \; ,\\
\nonumber
\alpha &=&  \frac{n_e}{5 \times 10^{-5} {\rm cm}^{-3}} \times
\frac{\la k T_e \ra}{ 2 \, {\rm keV}} \; .
\eea
This level of temperature distortion is consistent with the COBE 
FIRAS limit on deviations
from the blackbody spectrum on large angular scales \cite{FIRAS}.

Recently, Dolag et al. \cite{Dolag05} have studied the imprint
of the local superclusters on the CMB via the SZe. They have used
constrained simulations to study these local structures, and
their conclusion was that the thermal SZe is too small, by at
least one order of magnitude, to affect either the quadrupole 
or the octopole. It is clear that the SZe model 
presented here cannot account for the anomalous quadrupole 
and octopole if the density
and temperature of the gas in the ISC medium
lie near the conventionally accepted limits, which are derived
in part from observations and in part from simulations such as
those done by Dolag et al., which include
baryons, gas flows and feedback mechanisms. Therefore, if our
model is correct, then either the mechanisms that 
endow the ISC medium with gas are still not entirely understood, or
the global parameters and initial conditions have to be changed 
(which is less likely.) Another possibility is that the SZe is
not the only source of foreground, in which case a
combination of foregrounds (all spatially correlated so the
effects add up) is responsible for the distortions. In fact, 
such a combination of effects is not unlikely, since all
local structures are composed of multi-phase gas and other
foreground sources.



\section{Geometry and location of the foreground}

In Table I we showed the effect of subtracting the HFg
based on the projected volume of the LSC. This HFg
peaks at the center of the LSC and falls steadily
as the line of sight moves from the center of the LSC.
When the HFg is subtracted from a full CMB sky, 
both the quadrupole is enhanced and the quadrupole-octopole alignment
is weakened. This means that, whatever the source of the HFg
in the LSC (or its specular image), the geometry of the LSC
is such that the foreground's phases for the quadrupole and octopole 
work in the direction of correcting for the low quadrupole and
for the high quadrupole-octopole alignment of the CMB maps.

We can test the spatial location, shape and orientation of 
this HFg,
and check whether these are
indeed correlated with the properties of the LSC, or if the 
improvement in the quadrupole level and in the alignments 
are just flukes that could have happened whatever the location
of the HFg.
This can be done by rotating the foreground maps by arbitrary
Euler angles, and then computing the effect of subtracting them
from the CMB maps.

We employ an unbiased measure of the likelihood of the maps
defined by:
\be
\label{estimator}
P_{Tot} = {2^4} P_+(C_2) P_-(C_2) P_+(S) P_-(S)
= {2^4} P_+(C_2)[1-P_+(C_2)] P_+(S) [1- P_+(S)] \; .
\ee
This estimator is maximal ($P_{Tot}=1$) 
for a map whose quadrupole $C_2$ and alignment $S$ are equal to their 
respective expectation values -- in which case 
$P_-(\bar{C}_2) = P_+(\bar{C}_2) = P_-(\bar{S}) = P_+(\bar{S}) = 1/2$.

We will consider two HFg models:
the LSC model described by Eq. (\ref{ra}) (hereafter HFg-LSC), 
and a model inspired by the proposal by
Inoue \& Silk \cite{Disk06} --
a nearly homogeneous disk with a diameter of $50^o$ (hereafter HFg-D.)
It should be noticed that the HFg-D
model must have a higher 
temperature distortion in order to cause the
same order-of-magnitude effect in the likelihoods, compared to the 
HFg-LSC model. The reason is simple, and can be inferred from Fig. 
4: since the HFg-D model 
has nearly homogeneous temperature, all low multipoles have similar amplitudes. 
The temperature in the HFg-LSC model, on the other hand, falls steadily from
the center of the LSC, and this angle dependence coincides roughly with
the angular dependence of the quadrupole. 
Therefore, the amplitude of the temperature distortion in the HFg-D model 
must be higher than in the HFg-LSC model in order to get the same
level of the quadrupole. So, whereas in the HFg-LSC model the peak temperature
distortion is of order $-90 \mu$K, with a quadrupole $C_2 \approx 60 \mu$K$^2$,
for the HFg-D model to get the same quadrupole the temperature
of the $50^o \times 50^o$ spot must be $T_{Disk} \approx - 130 \mu$K.

For each model we rotate the putative foreground by random Euler angles and 
subtract it from a CMB map to obtain the ``cleaned'' map. We then look for 
the corrected maps with highest likelihoods, as measured by the largest
value of the unbiased probability $P_{Tot}$.
We used 2000 random rotations for each foreground model and for each 
map of Table I (all with the KP2 mask applied.)

The results are shown in Figs. 5-16, where each spot marks the
location of a hypothetical foreground. The dark spots mark the locations
of the top 5\% of foregrounds as measured by the likelihood of the 
cleaned maps. The lighter, smaller spots mark the location of the top
10\%, 20\% and so forth. The bottom 50\% locations are shown as the 
smallest, lightest spots. The top foreground in each case is shown in 
Table II, along with its location and likelihood.

\begin{figure}
\includegraphics[width=8cm]{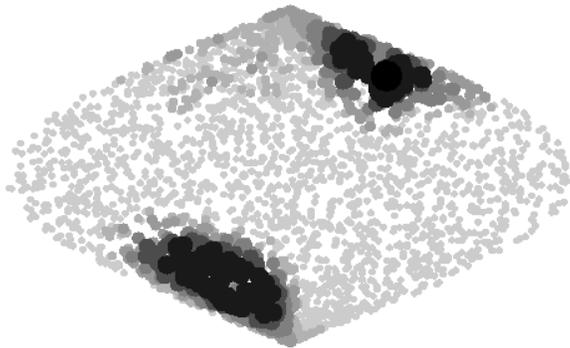}
\caption{\label{fig:5} Locations of the HFg-LSC
model applied to the ``mask 0'' TOH map \cite{TOH,OT06}.
The largest, darkest spot marks the top foreground 
as measured by the highest values of $P_{Tot}$ obtained after
3000 random rotations of the putative foreground.
The top 5\%, 10\%, 15\% and 33\% foregrounds 
are indicated by the progressively smaller and lighter spots.
The smallest, lightest spots mark the bottom 67\% locations. 
Here and in Figs. 6-16, instead of the Mollweide
projection, we use a simple map $(\theta,\varphi) \rightarrow (\theta,\varphi \, \sin{\theta} )$.}
\end{figure}

\begin{figure}
\includegraphics[width=8cm]{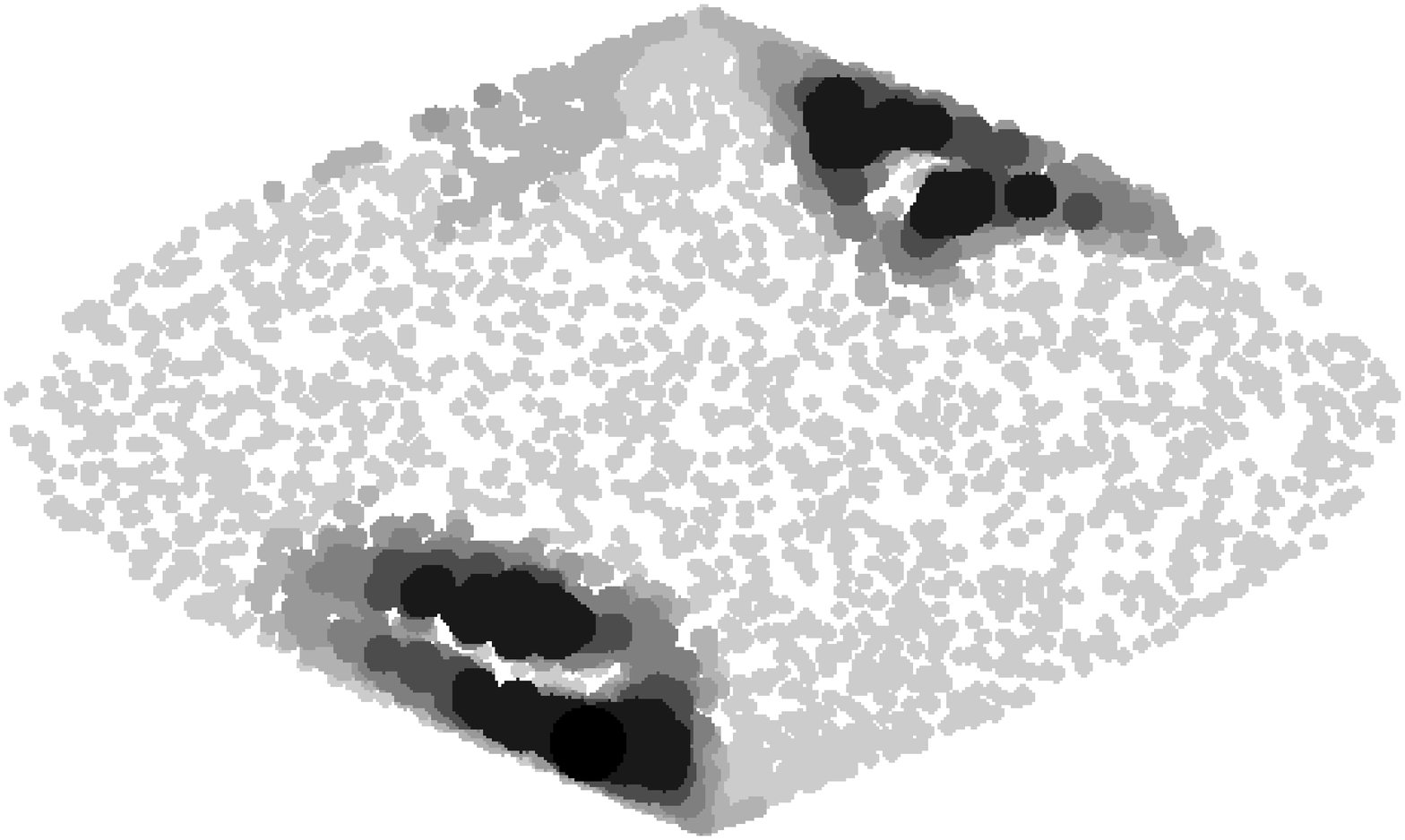}
\caption{\label{fig:6} Same as before, in the
case of the 
HFG-D model applied to the ``mask 0'' TOH map \cite{TOH,OT06}.}
\end{figure}

\begin{figure}
\includegraphics[width=8cm]{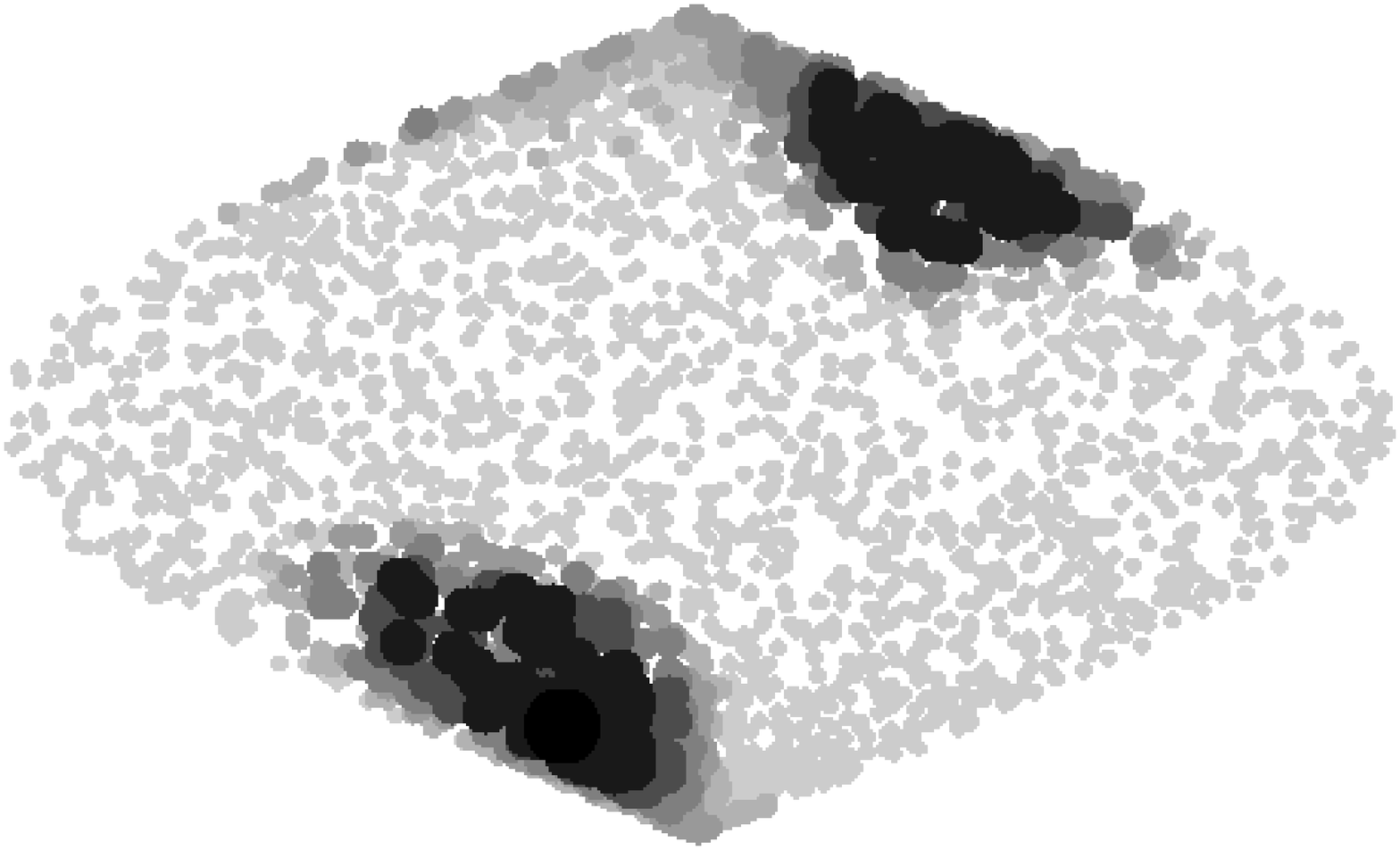}
\caption{\label{fig:7} Same as before, in the
case of the HFg-LSC
model applied to the ``mask 6'' TOH map \cite{TOH,OT06}.}
\end{figure}

\begin{figure}
\includegraphics[width=8cm]{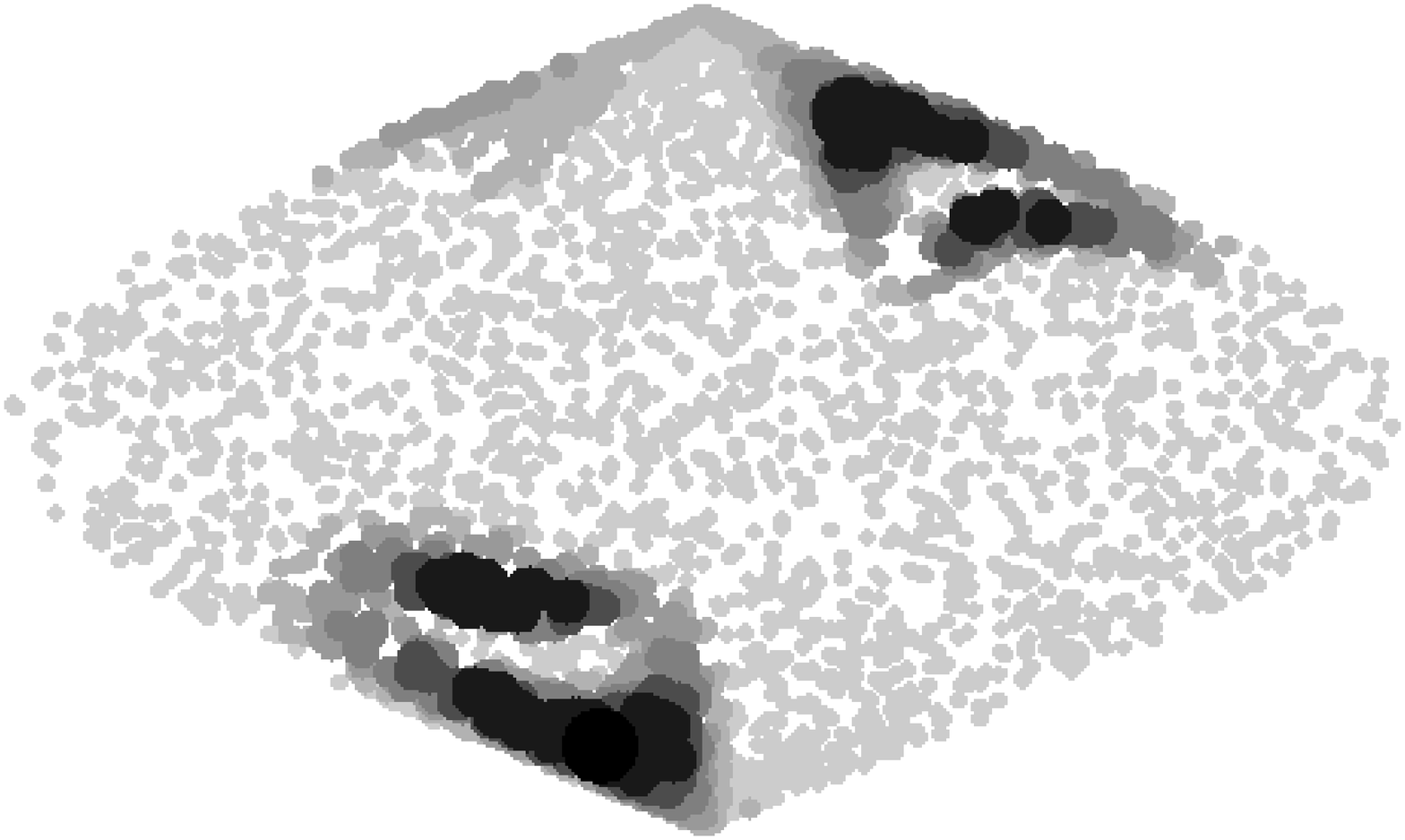}
\caption{\label{fig:8} Same as before, in the
case of the 
HFg-D model applied to the ``mask 6'' TOH map \cite{TOH,OT06}.}
\end{figure}

\begin{figure}
\includegraphics[width=8cm]{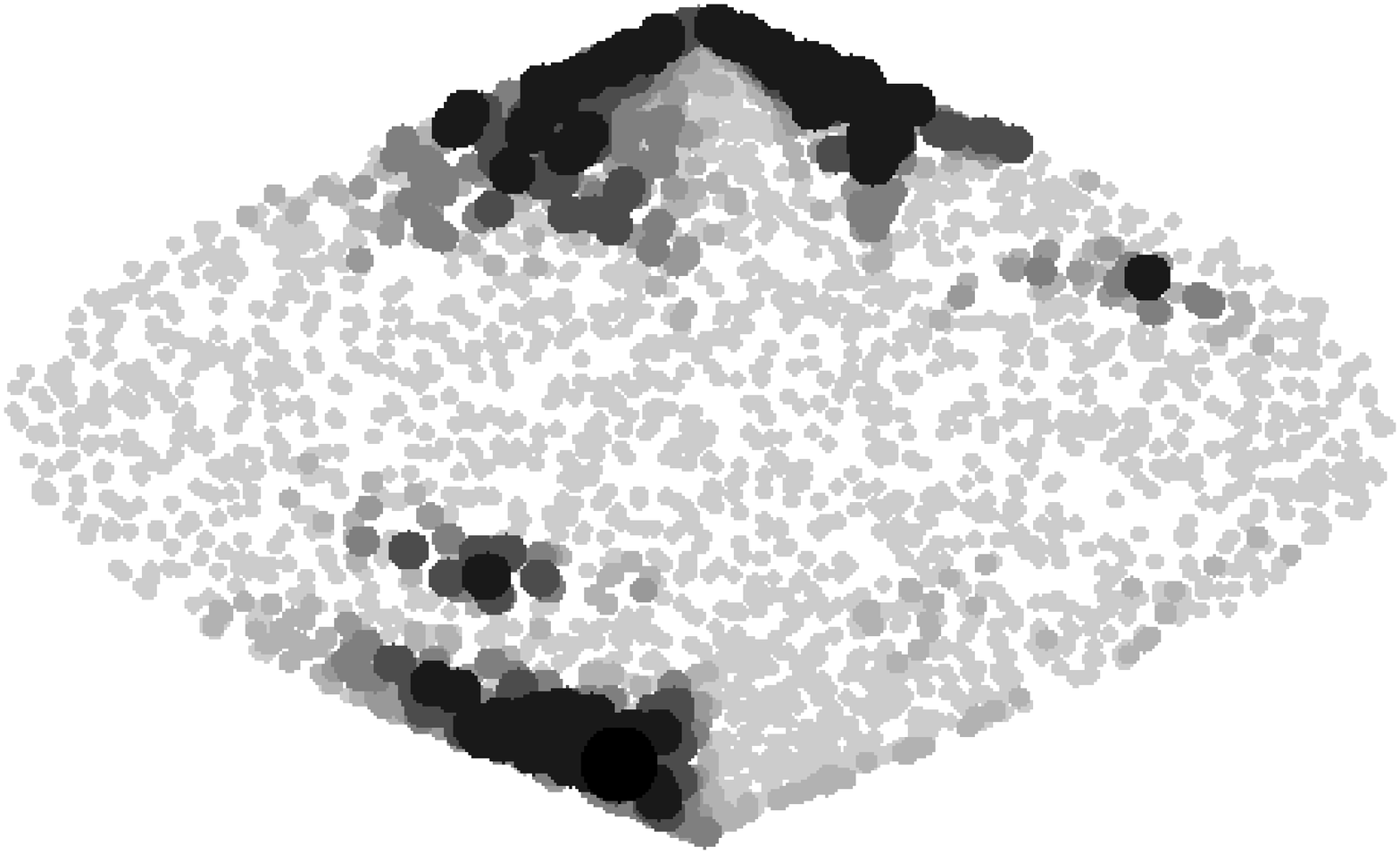}
\caption{\label{fig:9} Same as before, in the
case of the 
HFg-LSC model applied to the 1-year Coadded map.}
\end{figure}

\begin{figure}
\includegraphics[width=8cm]{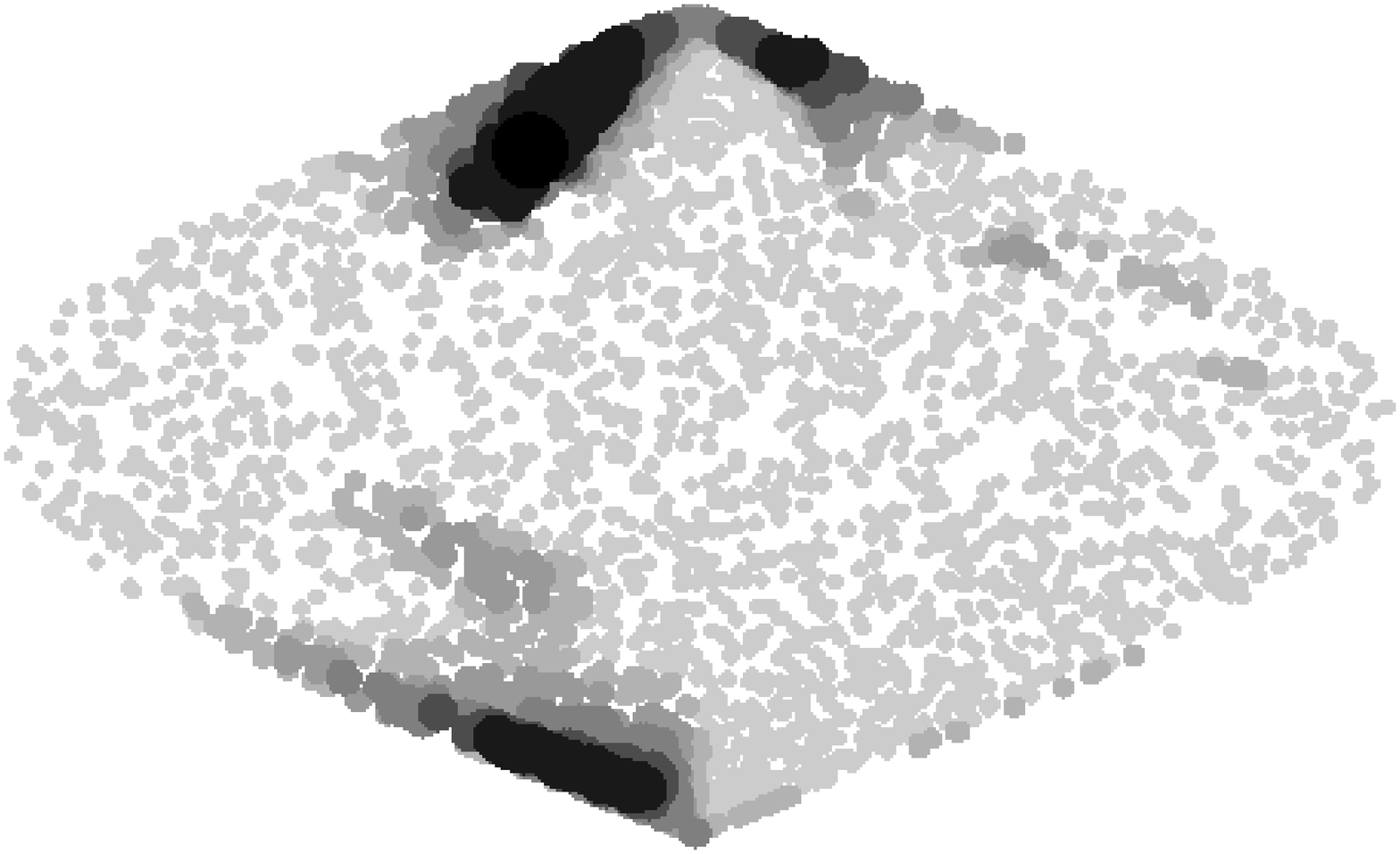}
\caption{\label{fig:10} Same as before, in the
case of the 
HFg-D model applied to the 1-year Coadded map.}
\end{figure}

\begin{figure}
\includegraphics[width=8cm]{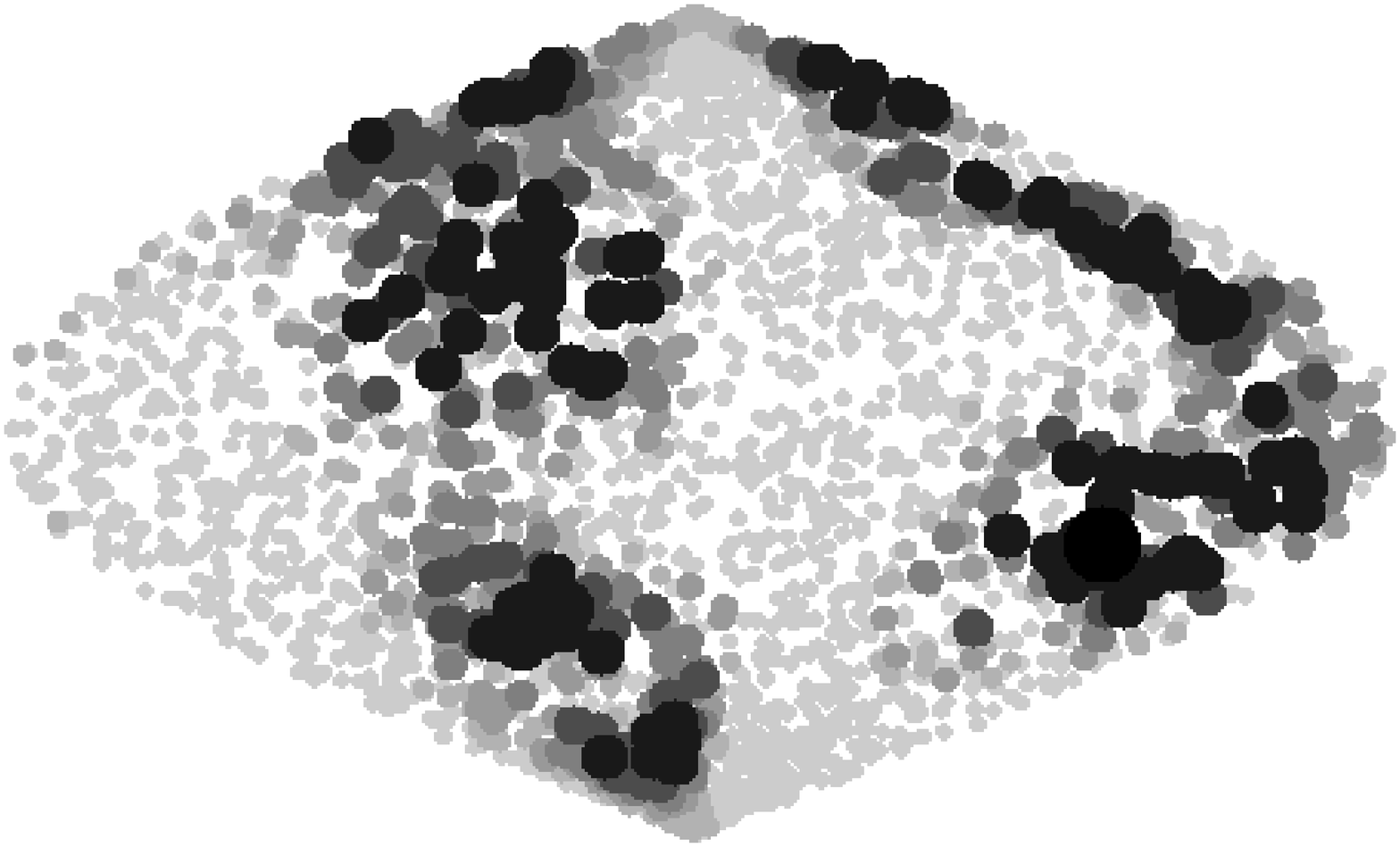}
\caption{\label{fig:11} Same as before, in the
case of the 
HFg-LSC model applied to the 3-year Coadded map.}
\end{figure}

\begin{figure}
\includegraphics[width=8cm]{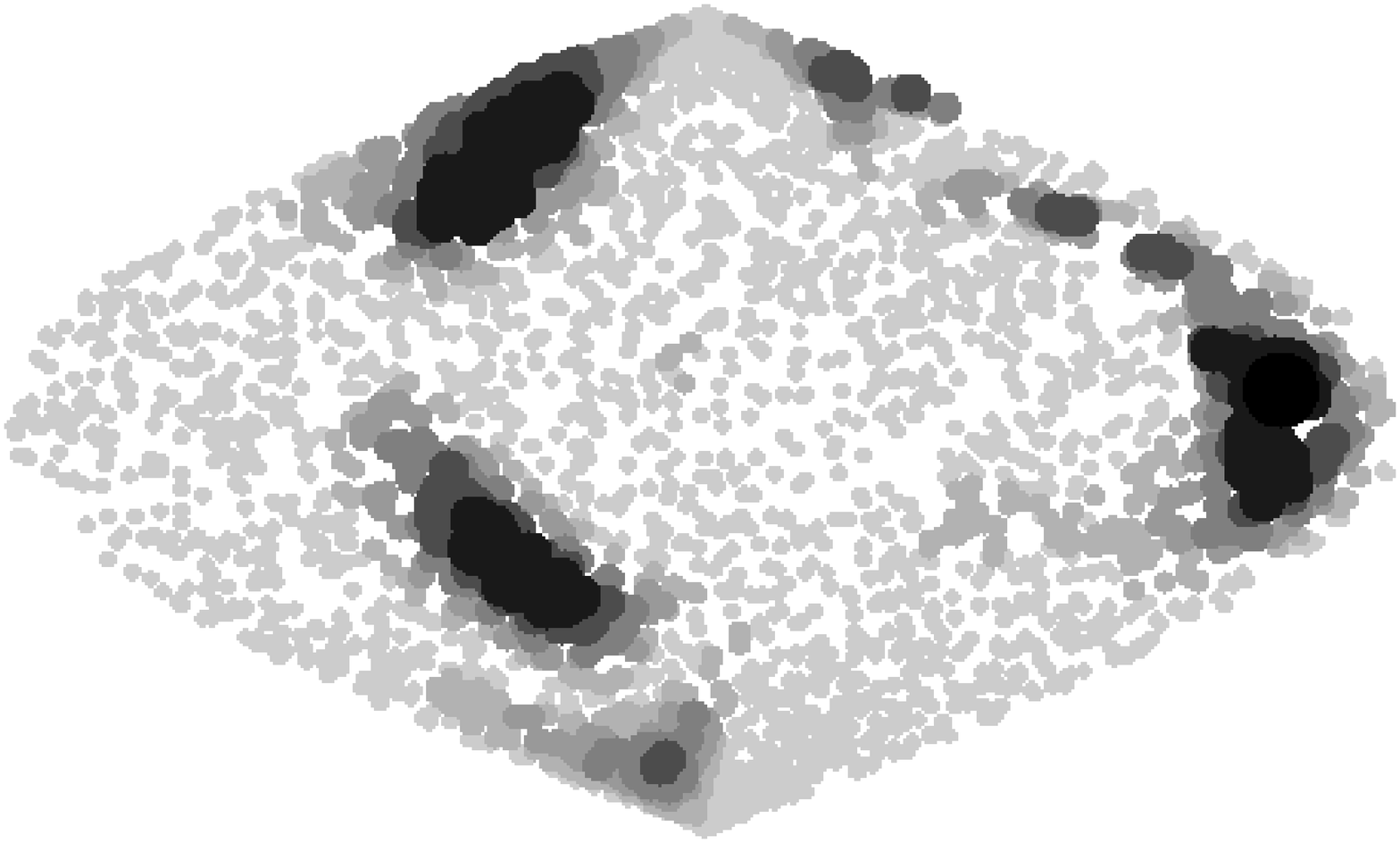}
\caption{\label{fig:12} Same as before, in the
case of the 
HFg-D model applied to the 3-year Coadded map.}
\end{figure}

\begin{figure}
\includegraphics[width=8cm]{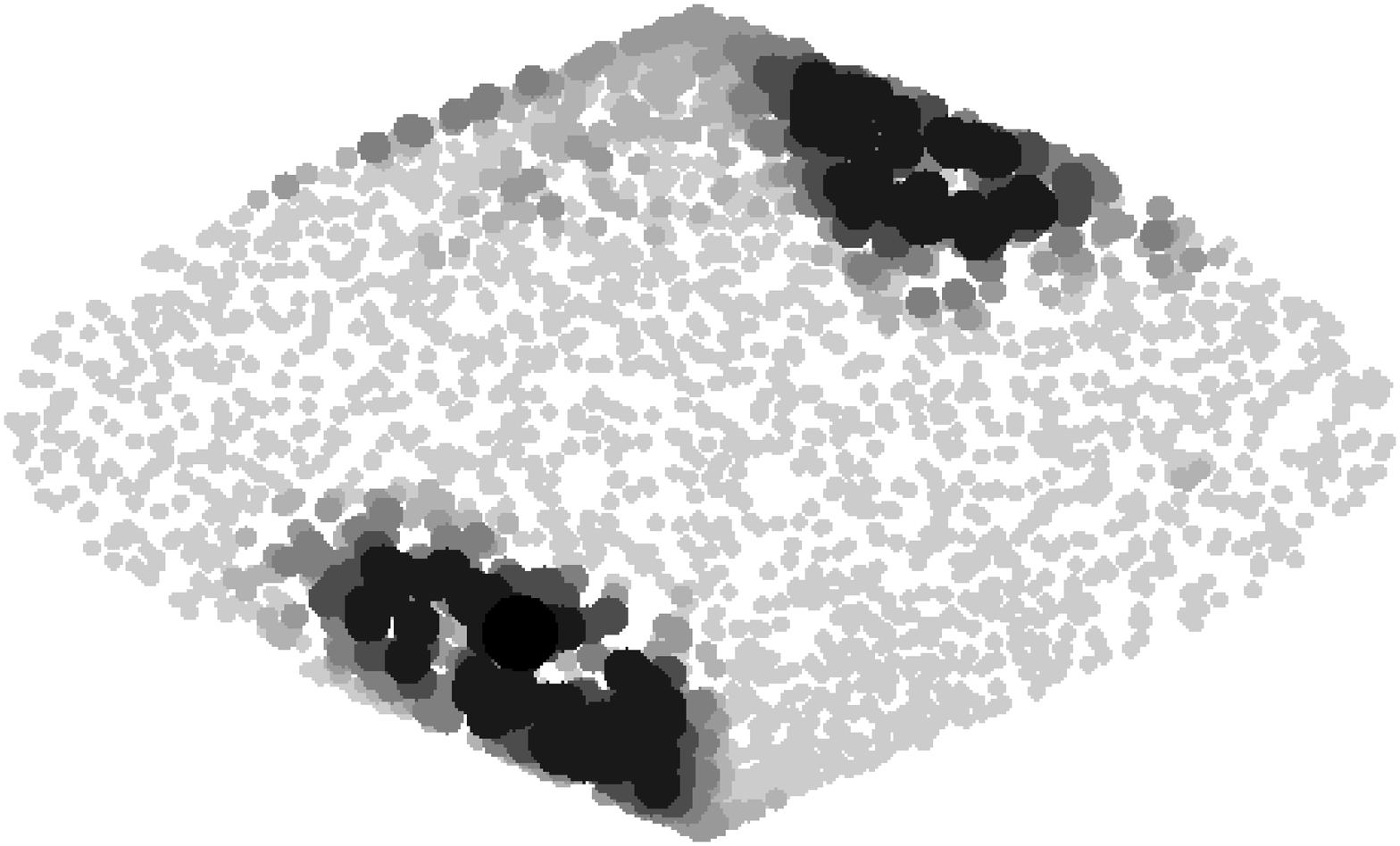}
\caption{\label{fig:13} Same as before, in the
case of the 
HFg-LSC model applied to the 1-year ILC map.}
\end{figure}

\begin{figure}
\includegraphics[width=8cm]{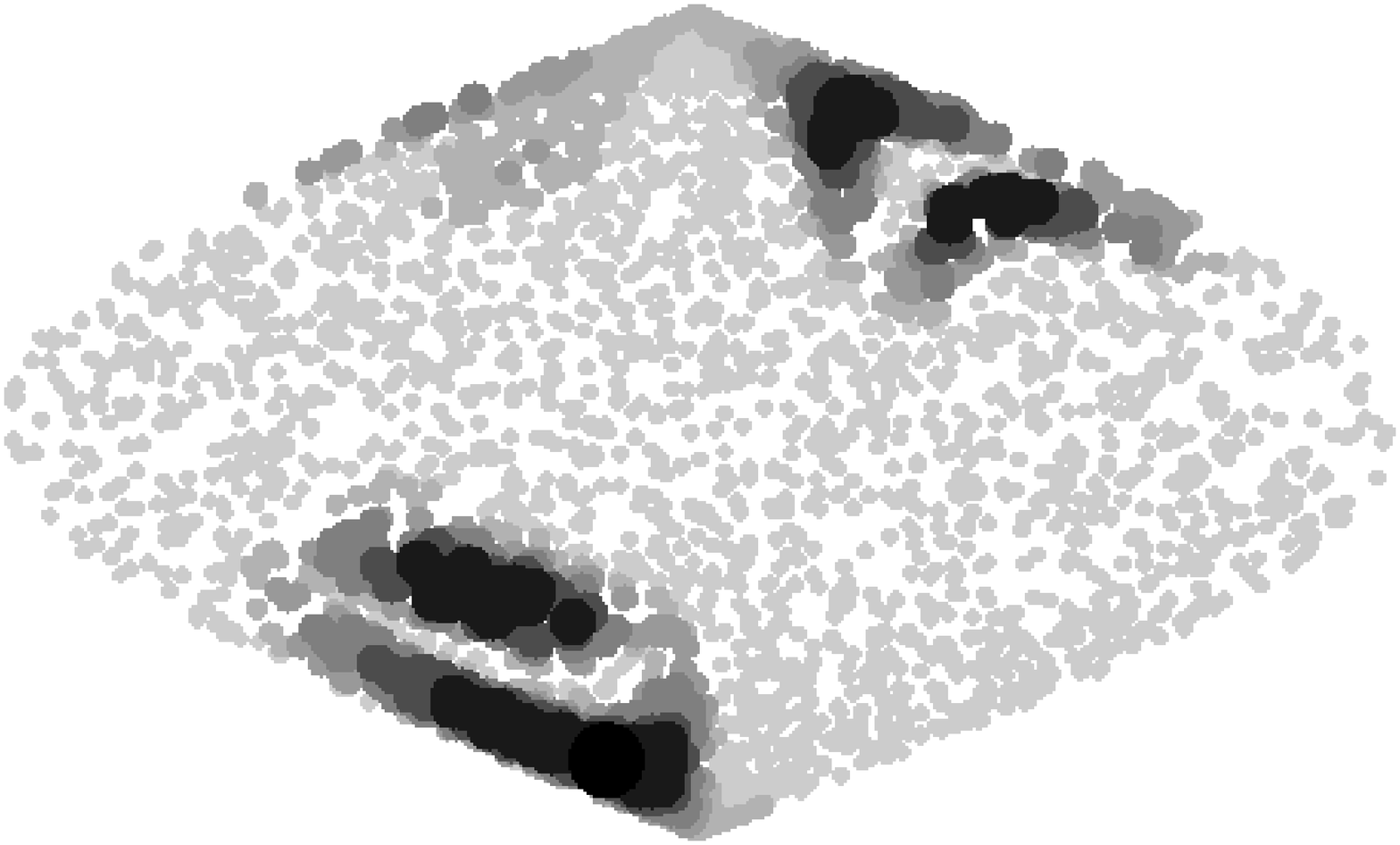}
\caption{\label{fig:14} Same as before, in the
case of the 
HFg-D model applied to the 1-year ILC map.}
\end{figure}

\begin{figure}
\includegraphics[width=8cm]{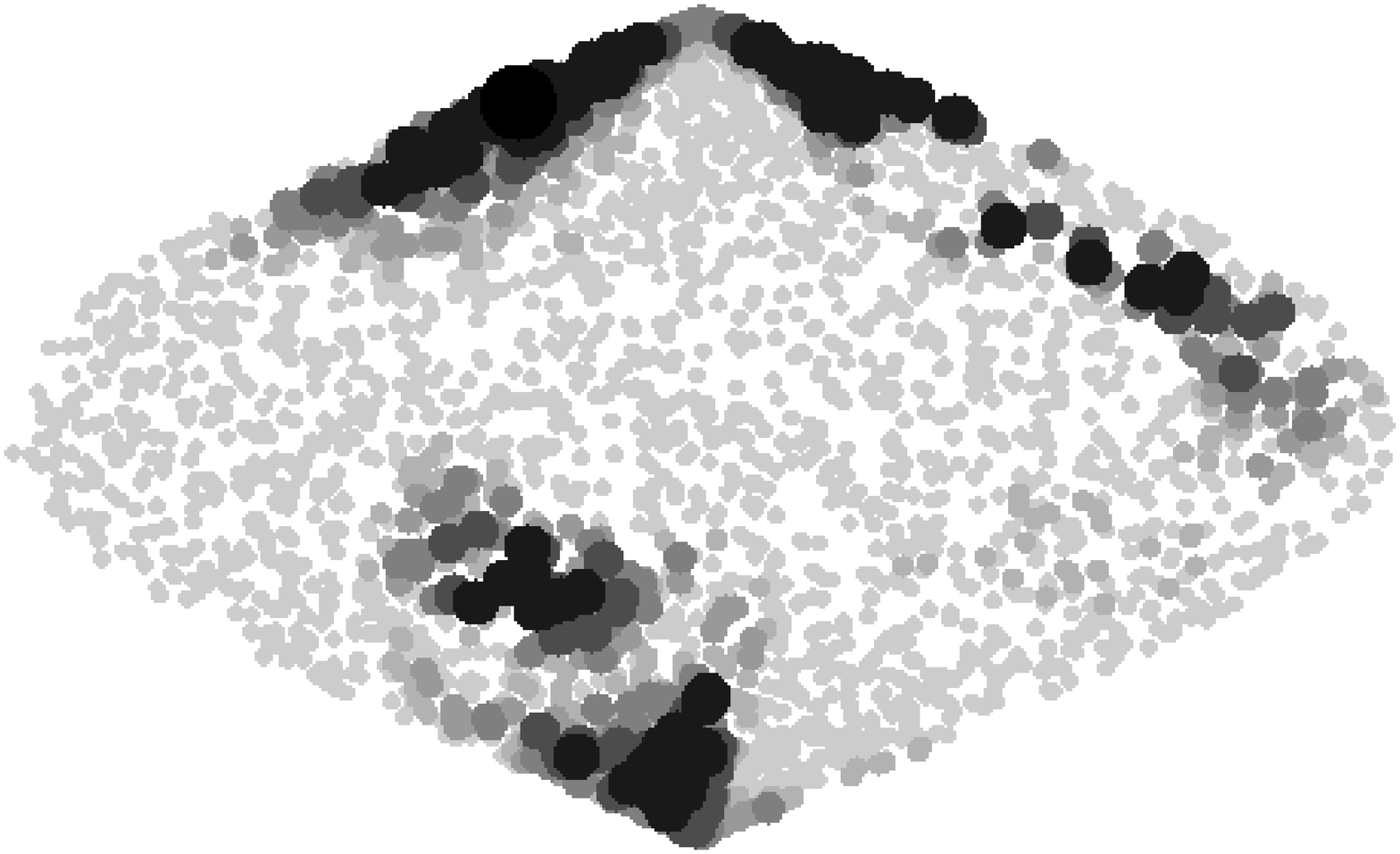}
\caption{\label{fig:15} Same as before, in the
case of the 
HFg-LSC model applied to the 3-year ILC map.}
\end{figure}

\begin{figure}
\includegraphics[width=8cm]{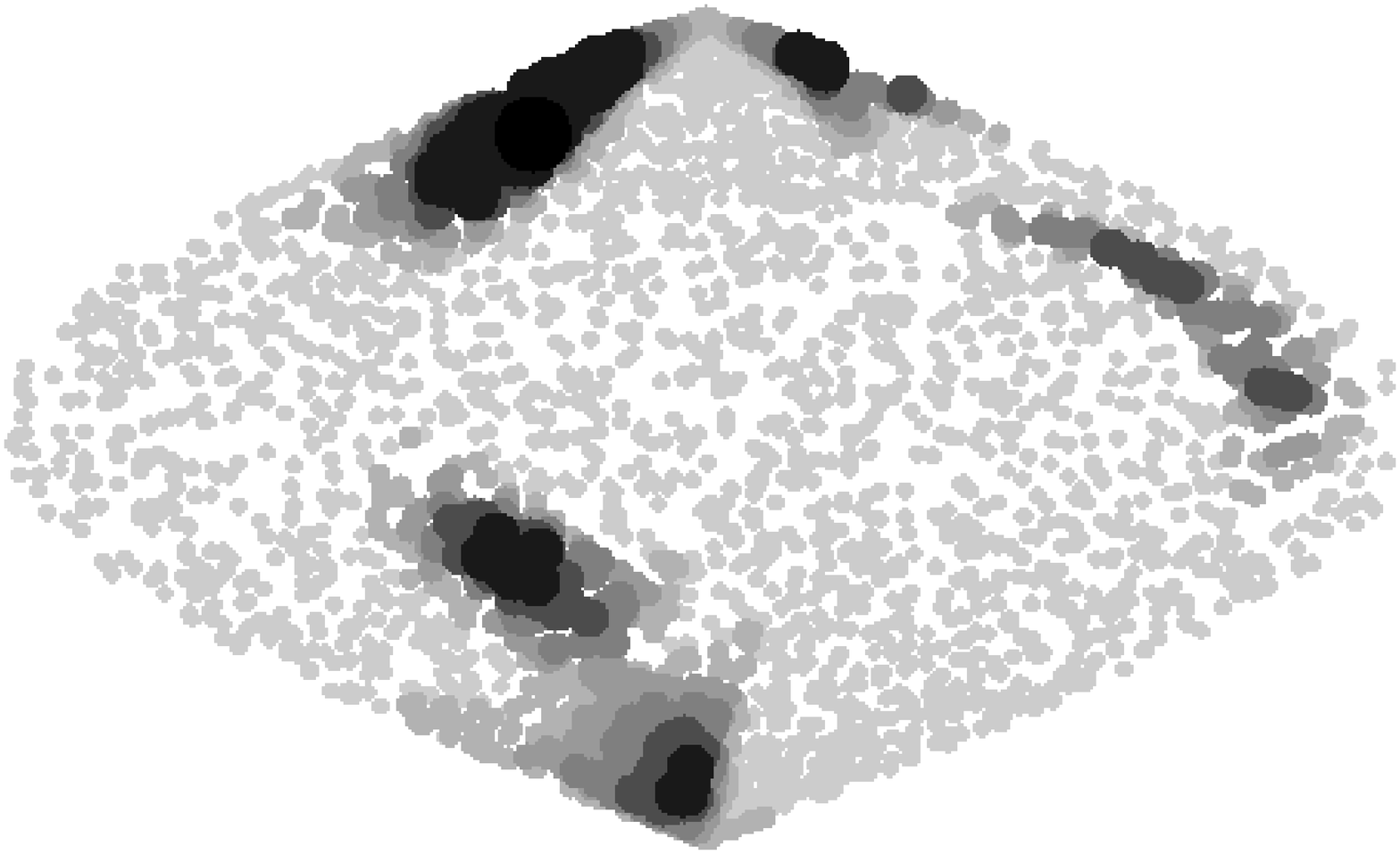}
\caption{\label{fig:16} Same as before, in the
case of the 
HFg-D model applied to the 3-year ILC map.}
\end{figure}

\begin{table}
\begin{tabular}{|c|c|c|c|}
\hline
CMB Map - Foreground & $\quad P_{tot} \quad $ & $\quad (l,b) \quad$ \\
\hline
\hline
TOH$_{\rm Mask \; 0}$ - LSC & 0.380 & (136,-51)\\
TOH$_{\rm Mask \; 0}$ - Disk  & 0.231 & (152,-69) \\
TOH$_{\rm Mask \; 6}$ - LSC & 0.488 & (156,-63)\\
TOH$_{\rm Mask \; 6}$ - Disk  & 0.363 & (159,-73) \\
Coadded (1yr) - LSC & 0.182 & (209,78)\\
Coadded (1yr) - Disk  & 0.144 & (137,50) \\
Coadded (3yr) - LSC & 0.155 & (314,-10)\\
Coadded (3yr) - Disk  & 0.108 & (122,50) \\
ILC (1yr) - LSC & 0.219 & (141,77)\\
ILC (1yr) - Disk & 0.222 & (144,65)\\
ILC (3yr) - LSC & 0.216 & (133,74)\\
ILC (3yr) - Disk & 0.155 & (129,61)\\
\hline
\end{tabular} 
\caption{\label{Table2} Rotated foregrounds with the highest likelihoods, 
obtained by maximizing $P_{Tot}$ in the cleaned maps. We 
have normalized the foregrounds by their quadrupoles, 
which we set to be $C_{2}^{Fgrd} \approx 60 \mu$K$^2$. In terms of the 
HFg-LSC model, the parameter $\alpha = 1$. In terms of the 
HFg-D model, the temperature
of the $50^o \times 50^o$ cold spot is $T_{Disk} \approx - 130 \mu$K.
}
\end{table}

It can be seen from Figs. 5-16 that, in most cases, there is a strong
clustering of the preferred locations around Virgo (and the LSC), 
its diametrically opposite point, and the region which correspons to 
Virgo rotated $180^o$ around the $z$-axis. The preferred locations 
seem to be more scattered for the 3-year Coadded map 
with both foreground models, and for the ILC maps with the 
HFg-LSC model.

There is a simple explanation for the spatial distribution seen
throughout Figs. 5-16: the 
HFg's change
$P_{Tot}$ mostly by amplifying and rotating the quadrupole
of the original maps. But the quadrupole is even under parity
transformations $\hat{n} \rightarrow -\hat{n}$, therefore from the
point of view of the quadrupole, it is irrelevant if the foreground
is at the LSC or at its diametrically opposite side. Moreover, since most
of the power of the quadrupole lies in its $\ell=2,m=0$ and $\ell=2, m=\pm 2$
components, a rotation by $180^o$ degrees effects very little change to it. 
Therefore, if the foreground which is distorting the CMB is indeed 
at the LSC, then its preferred locations in a blind search will degenerate to 
not only the vicinity of the LSC itself, at $(l,b)=(284^o,74^o)$, but also to 
its ``dual points" at $(l,b) \approx (104^o, -74^o)$ and  
$(l,b) \approx (104^o, 74^o)$. This is indeed what seems to happen for most 
maps -- see Table 2.

It is interesting to notice that, even though the amplitude of the 
octopole of the 
HFg-LSC model is subdominant, in the  
HFg-D model there is a substantial change in the octupole after the subtraction.
Although the impact of the octopole in 
$P_{Tot}$ is small, there are other statistical tests which are sensitive
to it -- in particular, the $S^{(4,4)}$ statistic of Copi {\it et al.} 
\cite{Copi04,Schwarz04} which tests for an alignment of the quadrupole and 
octopole with the ecliptic plane. The 3-year Coadded
and ILC maps have too high values of $S^{(4,4)}$ at $99.5\%$ C.L. After
subtracting the best-fit Disk foreground model, those values come down
to $84\%$ C.L. for the Coadded map and to $98\%$ C.L. for the ILC map.

\subsection{Higher multipoles}

If the HFg indeed exists, it may be possible to 
detect correlations in the higher $\ell$ components as well. However,
depending on the model,
the spectrum of the foreground can decay with large $\ell$, and 
the increasing number of phases means that these correlations will probably
difficult to detect. If the foreground is homogeneous (as in the Disk model),
the higher multipoles can also become important and their presence may
affect the alignments between, say, $\ell=2-5$.

It is trivial to generalize the tests of Eqs. (\ref{def:S})-(\ref{def:D})
for higher multipoles -- see, e.g., \cite{KW04,Nois}. If a given
foreground cures the quadrupole level and the quadrupole-octopole alignment
at the cost of introducing anomalous alignments between other multipoles, then
the overall likelihood of the resulting map should fall. On the other
hand, if the original CMB map has another anomalous alignment
which is relaxed because of the hypothetical foreground, then the
likelihood of the resulting map increases even further.

We have tested the maps of Table II for signs of anomalous 
alignments between the higher multipoles using an unbiased probability
with 38 tests, $P_{38}$ -- which includes the level of the quadrupole and the
quadrupole-octopole alignment \cite{Nois}. The results are mixed, and probably 
reveal intrinsic differences between the maps, masks and foreground models.
For example, whereas the original TOH map with the Mask 6 \cite{TOH,OT06} has
$P_{38} = 3.5 \times 10^{-15}$, subtraction of the best-fit HFg-LSC
model leads to an improvement to $P_{38} = 3,5 \times 10^{-10}$, while 
subtraction of the best-fit HFg-D
model improves it to 
$P_{38} = 2,4 \times 10^{-7}$.
For the Coadded map (3-year data) with the KP2 mask \cite{WMAP3y,WMAP1}, 
the original map has 
$P_{38} = 4.4 \times 10^{-13}$, and after subtracting the best-fit 
HFg-LSC model it only improves to $P_{38} = 1,1 \times 10^{-9}$, 
while subtraction of the best-fit 
HFg-D model leads to $P_{38} = 3,7 \times 10^{-8}$.
Notice that in all cases above, the level of the quadrupole and the 
quadrupole-octopole alignment alone are responsible for a factor of $10^{2}$ 
improvement in $P_{38}$ for the TOH map, and for a factor of $\sim 10$ 
improvement in $P_{38}$ for the ILC map.


\section{Conclusions}

We have presented circumstantial evidence that an extended 
foreground near the
dipole axis could be distorting the CMB. The subtraction of 
such a foreground increases the quadrupole, 
removes the
(anomalous) quadrupole-octopole alignment, and dramatically
increases the overall likelihood of the CMB maps. Possible
physical mechanisms that could account for this foreground
are the Sunyaev-Zeldovich effect \cite{AS} and the Rees-Sciama
effect \cite{ReesSciama}, although it should be noted that both
options only work in extreme situations that are probably 
unrealistic. Another possibility is that a combination of
effects is responsible for the foreground.
However, if the Sunyaev-Zeldovich effect due to
the LSC's gas is indeed responsible for the foreground, it 
could be directly observed by the Planck satellite 
\cite{PLANCK} within the next few years -- see also \cite{Dolag05}.

We have also shown that the phases of the CMB maps are such
that the optimal places for such foregrounds to exist would be around
the Local Supercluster, its specular image, or the site of the
Local Supercluster rotated $180^o$ around the galactic polar axis.
Furthermore, of the two foreground models analyzed here, the
non-uniform foreground (HFg-LSC model) seems preferred by the
data as it needs a lower overall temperature distortion
in order to improve the likelihood of the CMB maps.

\vskip 0.2cm

\noindent
We would like to thank S. Boughn, M. Coutinho, G. Holder, 
R. Rosenfeld, D. Schwarz, M. Tegmark, T. Villela and I. Waga 
for useful comments during various phases of this project.
This work was supported by FAPESP and CNPq.



\end{document}